\documentclass[11pt]{iopart}
\usepackage{graphicx, color,iopams}
\usepackage[pdftex, bookmarks, colorlinks=true, bookmarksnumbered=true, pdfstartview=FitH, pdfpagemode=UseOutlines, pdfstartpage=1,bookmarksopen=true]{hyperref}
\hypersetup{colorlinks,citecolor=blue,filecolor=blue,linkcolor=blue,urlcolor=blue, pdftex}

\begin{document}

\title[Hierarchical veto]{A hierarchical method for vetoing noise transients in gravitational-wave detectors}


\author{Joshua~R~Smith$^1$, Thomas Abbott$^1$, Eiichi Hirose$^{2,7}$, Nicolas Leroy$^3$, Duncan MacLeod$^4$, Jess McIver$^5$, Peter Saulson$^2$, Peter Shawhan$^6$}


\address{$^1$Department of Physics, California State University Fullerton, Fullerton, CA 92831, USA}
\address{$^2$Department of Physics, Syracuse University, Syracuse, NY 13244, USA}
\address{$^3$LAL, Universite Paris-Sud, IN2P3/CNRS, F-91898 Orsay, FR}
\address{$^4$Cardiff University, Cardiff, CF24 3AA, UK}
\address{$^5$University of Massachusetts - Amherst, Amherst, MA 01003, USA}
\address{$^6$University of Maryland, College Park, MD 20742, USA}
\address{$^7$Now at: Institude for Cosmic Ray Research, University of Tokyo, Tokyo, JP}

\ead{josmith@fullerton.edu}

\begin{abstract}
Non-Gaussian noise transients in interferometric gravitational-wave detectors increase the background in searches for short-duration and un-modelled signals. We describe a method for vetoing noise transients by ranking the statistical relationship between triggers in auxiliary channels that have negligible sensitivity to gravitational waves and putative gravitational-wave triggers in the detector output. The novelty of the algorithm lies in its hierarchical approach, which leads to a minimal set of veto conditions with high performance and low deadtime. After a given channel has been selected it is used to veto triggers from the detector output, then the algorithm selects a new channel that performs well on the remaining triggers and the process is repeated. This method has been demonstrated to reduce the background in searches for transient gravitational waves by the LIGO and Virgo collaborations.  
\end{abstract}

\section{Introduction}\label{secn:intro} 

The first generation of kilometer-scale interferometric gravitational-wave detectors, LIGO~\cite{ligo}, Virgo~\cite{virgo} and GEO\,600~\cite{geo}, have completed several years of network observation of the 40-10000 Hz frequency range. The data has been searched for various types of gravitational radiation including stochastic sources such as the early universe~\cite{stoch}, continuous sources such as spinning neutron stars~\cite{cw}, coalescence of binary systems of black holes or neutron stars~\cite{cbc}, and searches for un-modelled or poorly modelled bursts such as supernovae~\cite{burst}. So far no gravitational-wave detection has been made, however data analysis is ongoing. The next generation of detectors including Advanced LIGO~\cite{aligo} and Advanced Virgo~\cite{avirgo} are expected to observe gravitational waves from compact binary coalescence~\cite{rates} within this decade. 

Even with highly sensitive detectors, gravitational-wave searches are limited by noise. In addition data from all interferometric gravitational-wave detectors to date has shown a characteristic large non-Gaussian tail from non-astrophysical sources. In general, the shorter and less well-modelled a true signal is, the more difficult it is to distinguish from noise transients using signal processing. Requiring coincidence and coherence among multiple widely-separated detectors is an important and effective way to reduce the influence of transients. Still, the performance of searches for un-modelled bursts and high-mass binary coalescence signals  (which have short duration in these detector's frequency band) is greatly diminished by transients in the detector data.  This sets a practical limit on the sensitivity of the searches and on the false alarm rate that can be ascribed to candidate gravitational-wave signals. 

Interferometric gravitational-wave detectors are designed to be isolated from all significant non-gravitational-wave external phenomena (seismic, electromagnetic, acoustic), and they are equipped with systems to monitor both the local environment and auxiliary interferometer channels for disturbances. In addition, there is a large effort to identify poor quality data and to link these to causes in the local environment or to aspects of the instrument itself~\cite{nelson, slutsky, glitchgroup} so that the noise transients can be removed through improvements to the instrument or by ``vetoing''~\cite{upv, dicredico, ajith}, whereby periods of demonstrated low-quality data are removed from an analysis. The method described in this paper and the ``used percentage veto'' described in~\cite{upv} were the two methods most extensively used during the most recent science runs for both LIGO and Virgo. 

\begin{figure}[ht]
\centering
\includegraphics[scale=0.71]{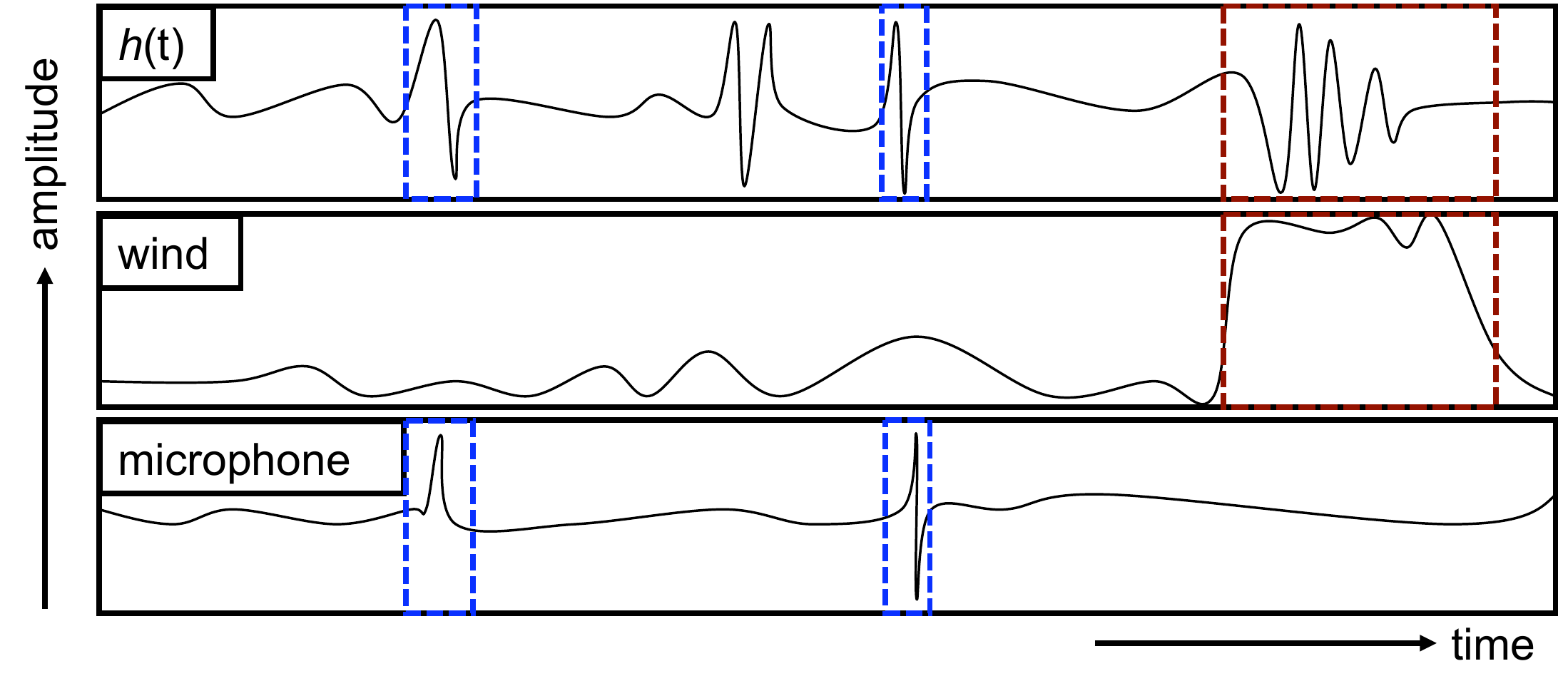}
\caption{Illustration of the removal of some data from the \emph{h(t)} channel due to its association with two hypothetical non-astrophysical disturbances, to obtain an improved data stream. The top trace, \emph{h(t)}, represents the \emph{h(t)} data. The middle trace is a monitor of wind speeds on the detector site, while the lowest trace is a microphone located in one of the detector's buildings. The first and second vetoed period in \emph{h(t)}, between pairs of dashed lines, are removed due to association with sharp glitches in the microphone, while the third period is removed because of high local wind speeds. This data removal would be done after a relationship between these types of disturbances and noise transients in \emph{h(t)} had been established.}
\label{fig:cartoon}
\end{figure}

Figure~\ref{fig:cartoon} shows a cartoon example of strain data in the detector output channel, referred to hereafter as \emph{h(t)}. Here some periods of \emph{h(t)} data are discarded because they are associated with hypothetical disturbances: high local wind speed and loud acoustic transients in a detector building. One useful indication of the effectiveness of a given veto is the ratio of the \emph{efficiency}, the percent of noise transients vetoed from \emph{h(t)}, to the \emph{deadtime}, the percent of the analysis time that is removed. If this ratio is high, the veto is considered useful. If the ratio is close to one, the veto performs no better than time removed at random.  In this paper we refer to noise transients that are deemed significant (i.e. by crossing some threshold) as \emph{triggers}. In the hypothetical situation shown in Figure~\ref{fig:cartoon}, three of four triggers in \emph{h(t)} are removed at a cost of about 25\% deadtime, giving a ratio of 3. Another indicator of veto utility is the percentage of an auxiliary channel's triggers that are coincident with a trigger in \emph{h(t)}, the so-called \emph{use percentage}. In Figure~\ref{fig:cartoon} the two triggers in the microphone channel each veto a trigger in \emph{h(t)}, so its use percentage is 100\%. 

In practice, auxiliary channels show excursions with a continuum of magnitudes, and there may be short time offsets between the disturbances and the triggers in \emph{h(t)}. Without \emph{a priori} knowledge of what are the relevant amplitudes and time windows to use, optimizing the many parameters needed to define an effective set of vetoes is a challenging problem. 

This paper describes a hierarchical veto algorithm called \emph{hveto}, used for the identification and removal of noise transients in searches for short-duration or poorly modelled gravitational waves. Its implementation for the LIGO and Virgo gravitational-wave detectors makes use of the hundreds of auxiliary channels recorded by each. It identifies the subset of channels that have negligible sensitivity to gravitational-wave signals, then determines which of these exhibit a significant relationship with transient noise present in \emph{h(t)}. Auxiliary channels are ranked based on their statistical significance, which quantifies how unlikely the number of time coincidences between triggers in \emph{h(t)} and triggers in auxiliary channel are, in comparison with the number expected by chance based on Poisson statistics. 

The most novel feature of hveto is that it is hierarchical. The basic idea of using noise transients detected in auxiliary channels to veto putative signals has been around since the prototype era. One shortcoming of these approaches was that the significance of each channel was evaluated individually with respect to \emph{h(t)}, leading to many channels being adopted as vetoes even though they were largely vetoing the same set of triggers. Since not all channels have high use percentage, this led to unnecessarily high deadtime. The goal of the hierarchical method is to find a minimal set of veto conditions that collectively has a high efficiency and low deadtime by selecting the best available veto, removing its effects from \emph{h(t)}, and then iterating the process on the remaining triggers to produce only as many vetoes as have a statistically significant effect.

Section~\ref{sec:sign} describes the statistic used by hveto to rank potential veto channels. Section~\ref{sec:algo} presents a flowchart and description of the hveto algorithm, and Section~\ref{sec:results} goes through a set of illustrative results from one week of LIGO data.  


\section{Statistical significance}\label{sec:sign}

\begin{figure}[ht]
\centering
\includegraphics[scale=0.71]{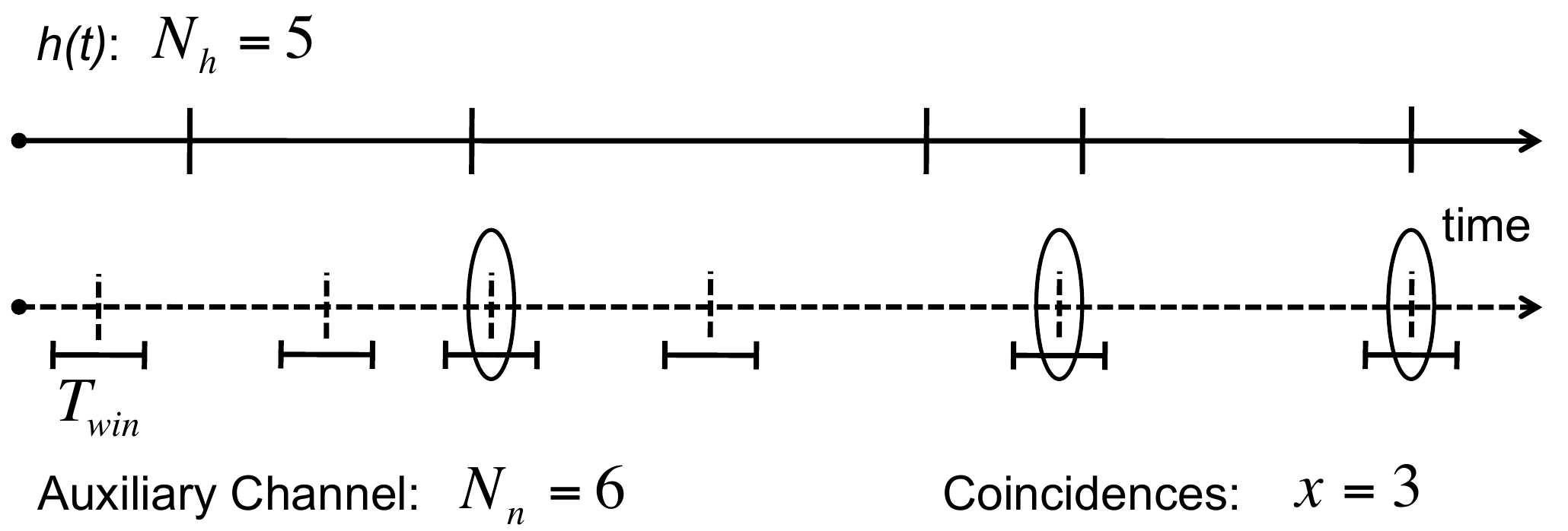}
\caption{Illustration of how coincidence is determined. Triggers in \emph{h(t)} are shown in the top trace. Triggers in the auxiliary channel, shown in the bottom trace, are assigned a time window. If an \emph{h(t)} trigger falls within the time window around an auxiliary channel trigger, the triggers are considered coincident, here coincident auxiliary channel triggers are circled.}
\label{fig:coincidence}
\end{figure}

The hierarchical veto algorithm looks for coincidence between two different types of triggers: putative gravitational-wave triggers from \emph{h(t)} and non-astrophysical triggers from auxiliary channels. The auxiliary channels are ranked based on the \emph{significance} of their relationship with \emph{h(t)}. First the number of time-coincidences between the triggers in a given channel and the triggers in \emph{h(t)} are counted, as shown in Figure~\ref{fig:coincidence}. Then the channels are ranked by a measure of how unlikely it is that their observed number of coincidences arose from the intersection of two random Poissonian time distributions with the same numbers of occurrences and time window. For each auxiliary channel, the significance is computed as\footnote{The value of $P_{poi} (\mu, k)$ is difficult to compute because its numerator is the factorial of a potentially large number. Alternatively the value can be computed using the incomplete gamma function. In our Matlab implementation, significance is calculated as, \texttt{sig(n) = -log10(gammainc(mu,k,'lower'))}. When $k$ is very large compared to $\mu$ the incomplete gamma function exceeds double precision limits and Matlab returns zero, resulting in infinite calculated $S$. In this case, we substitute a non-divergent approximation, $\sum_{k}^\infty{P_{poi}(\mu,k)} \approx P(\mu,k)$, which is implemented in Matlab as a sum of logarithms, \texttt{sig(n) = -k*log10(mu) + mu*log10(exp(1)) + gammaln(k+1)/log(10)}.},
\begin{equation}\label{eqn:sig}
S=-\log_{10}\left(\sum_{k=n}^{\infty}P(\mu, k)\right),
\end{equation}
where $n$ is the number of coincidences found between noise transients in that channel and noise transients in \emph{h(t)} during a given time $ T_{tot} $, and $ P(\mu ,k) $ is the Poisson probability distribution function,
\begin{equation}\label{eq:mu}
P (\mu, k) = \frac{\mu^{k} e^{-\mu}}{k!}.
\end{equation}
Here $\mu$ is the expected number of chance coincidences between random triggers in \emph{h(t)} and in the auxiliary channel, estimated by,
\begin{equation}
\mu = \frac{N_{h} N_{aux} T_{win}}{T_{tot}},
\end{equation}
where $ N_{h} $ and $ N_{aux} $ are the number of triggers in \emph{h(t)} and auxiliary channel, respectively, and $ T_{win} $ is the full width of the coincidence window used.

Shortly put, significance is $ -\log_{10} $ of the total probability of observing as many or more coincidences between two series of random occurrences than were actually observed~\footnote{A short numerical example is as follows. Consider a week of data, $T_{tot}=604800$ s, during which the number of \emph{h(t)} triggers is $N_{h}=1400$ and the number of triggers in a particular auxiliary channel is $N_{aux}=1100$. If a time window of $T_{win}=0.10$ s is used to check for coincidence, the expected number of coincidences is $\mu=0.25$. If the number of observed coincidences were one, the significance would be $S=0.65$, and if instead the number of coincidences observed were 30, the significance would be $S=50$.}  



\section{The hveto algorithm}\label{sec:algo}

\begin{figure}[ht]
\centering
\includegraphics[scale=0.75]{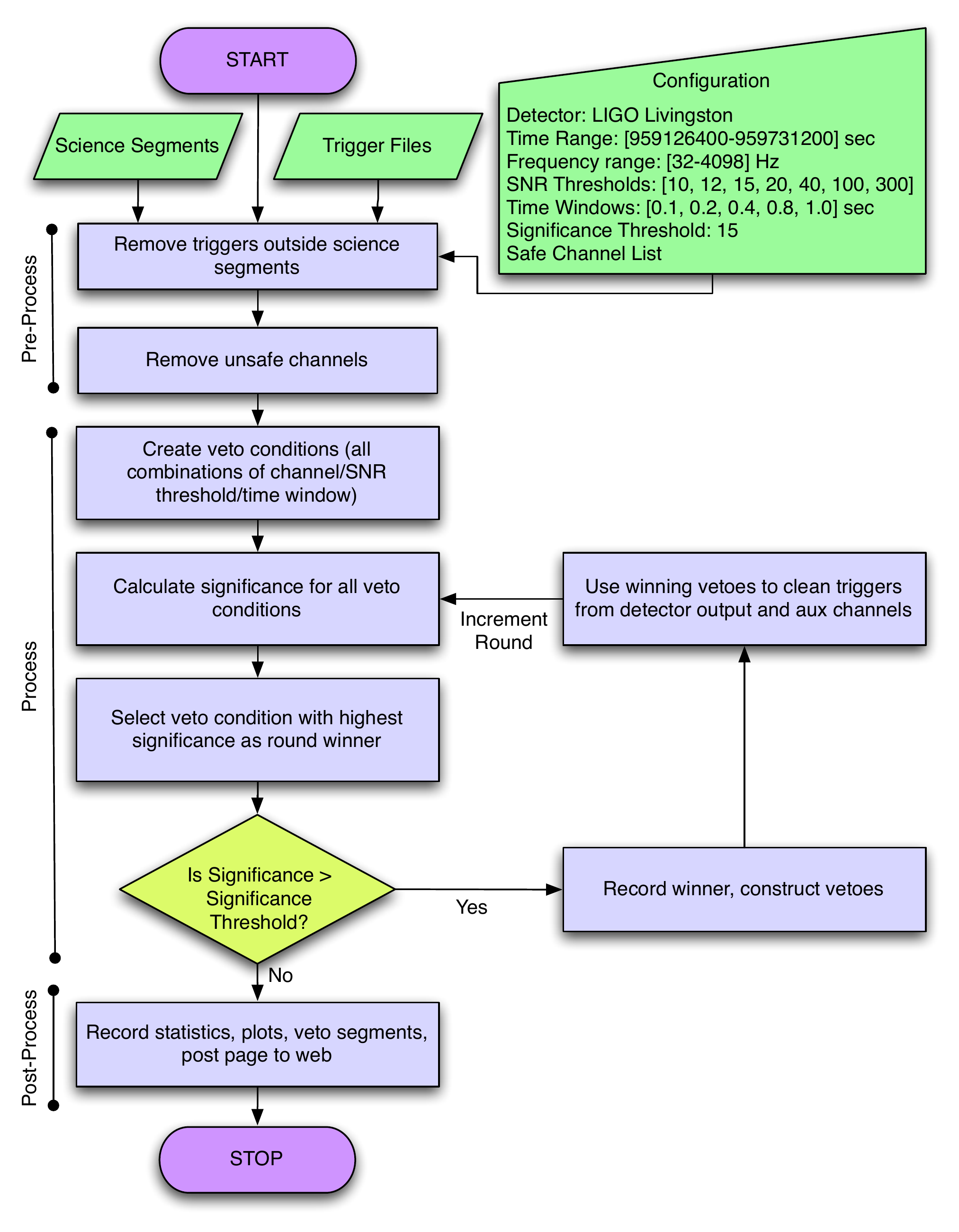}
\caption{Flowchart of the hveto algorithm described in the text. The configuration values are those used for Section~\ref{sec:results}.}
\label{fig:flowchart}
\end{figure}

A flowchart of the hveto algorithm is shown in Figure~\ref{fig:flowchart}. In the pre-processing stage, inputs consisting of a \emph{configuration file}, a list of \emph{science segments}, and a set of \emph{trigger files} are assembled. The configuration file contains all user-defined variables, including a detector name, a start and stop time defining the analysis period, a frequency range to consider, a list of thresholds on the signal-to-noise ratio (SNR) of triggers, a list of time-windows over which to check for coincidence, and a \emph{Significance Threshold}\footnote{The rate of noise transients in gravitational-wave data typically varies with time as conditions change. Therefore the mean rate of chance coincidences calculated from equation~\ref{eq:mu} may not be accurate. To account for this and other errors, a significance threshold can be determined empirically by using time-slide analysis (e.g., adding a seconds-long artificial offset to the central times of \emph{h(t)} triggers) to calculate the significance typical of the coincidences between auxiliary channels and \emph{h(t)} when there is no causal relationship. Significance of up to 5 is often observed.} below which the search for further vetoes will terminate. The list of science segments defines the subset of the full analysis period when the detector was operating normally. The trigger files contain a list of noise transients parametrized by their, time, frequency, and SNR. There is one trigger file for \emph{h(t)} and one for each auxiliary channel. Trigger files are produced beforehand by other algorithms, such as \emph{Kleinewelle}, \emph{Omega} (formerly Q-pipeline)~\cite{kwomega} and \emph{iHope}~\cite{ihope}, that identify short-duration candidate signals based on their excess power and/or likeness to signal models.

Vetoing \emph{h(t)} triggers based on a statistical relationship with auxiliary channels is only permissible if the auxiliary channels have negligible sensitivity to gravitational waves. Such channels are referred to as \emph{safe} for use in defining vetoes. Some auxiliary channels, particularly those measuring degrees of freedom of the interferometer, are weakly coupled to \emph{h(t)} and thus demonstrate a non-negligible response to gravitational-wave strain. During data taking runs, signals that mimic the behavior of gravitational waves are injected into the instruments by physically moving the test masses to make a time-varying strain signal. Many of these \emph{hardware injections} are performed at intentionally high SNR in the detectors to verify the safety of auxiliary channels; for a channel to be considered safe, it should not respond to hardware injections. To evaluate this, prior to the process outlined in Figure~\ref{fig:flowchart}, hveto compares a list of hardware injection times to auxiliary channel triggers, counts the number of coincidences within a 100 ms time window, and computes the significance of the coincidences in each channel. Any channel that is assigned a significance of greater than $S$=3 (conservative) is considered unsafe for use in veto production and removed~\footnote{This value of significance was chosen empirically to reject channels with known low-level coupling, without too many false rejections.}.

The Process stage of hveto proceeds in rounds. In the first round, a set of all possible \emph{veto conditions} (all combinations of auxiliary channels, time windows, and SNR thresholds given in the configuration) is created. Their times are compared to those in the \emph{h(t)} channel, coincidences are counted as described above, and the statistical significance of each channel is computed following Equation~\ref{eqn:sig}. The one channel/window/SNR combination with the highest significance is declared the winner of the first round.  

If the significance of the round winner is greater than the Significance Threshold, its trigger times are turned into veto segments by adding and subtracting half the time window width. The first round ends when all of the winner's veto segments are applied, removing those times from the segments over which \emph{h(t)} and all auxiliary channels are further analyzed.

The second round proceeds exactly as the first, by evaluating the statistical significance of each veto condition on the remaining \emph{h(t)} triggers and determining a winner. However, now the noise transients related to the winning channel from the first round have been vetoed. So the winner of the second round will be different than that of the first, and is likely to relate to a different noise mechanism. This is the key to the hierarchical operation of hveto. 

The rounds continue until the significance of a round-winning combination does not exceed the Significance Threshold. The result of the entire process is a short list of veto conditions, usually less than a dozen, that collectively have a high efficiency and low deadtime. 

Finally, in the Post-processing stage  the cumulative effect of all of the veto segments produced is assessed, statistics and plots (some of which are presented in the next section) are generated, and all of this information is posted to a web page.

\section{Illustrative results}\label{sec:results}


This section illustrates the utility of the hveto algorithm through application to one week of data from the LIGO Livingston Observatory between May 29 and June 4 2010. For more information on the LIGO systems mentioned here, see~\cite{ligo}. The \emph{h(t)} and auxiliary channel triggers were generated with Kleinewelle. The algorithm completed a total of 11 rounds before reaching the Significance Threshold of 15. The results indicated a variety of glitch mechanisms present in the detector. Two of the round winners, as well as the cumulative effects of the vetoes, are described below.  

\begin{figure}[ht]
	\begin{minipage}[t]{0.5\textwidth}
		\vspace{0pt}
		\includegraphics[width=1.2\linewidth]{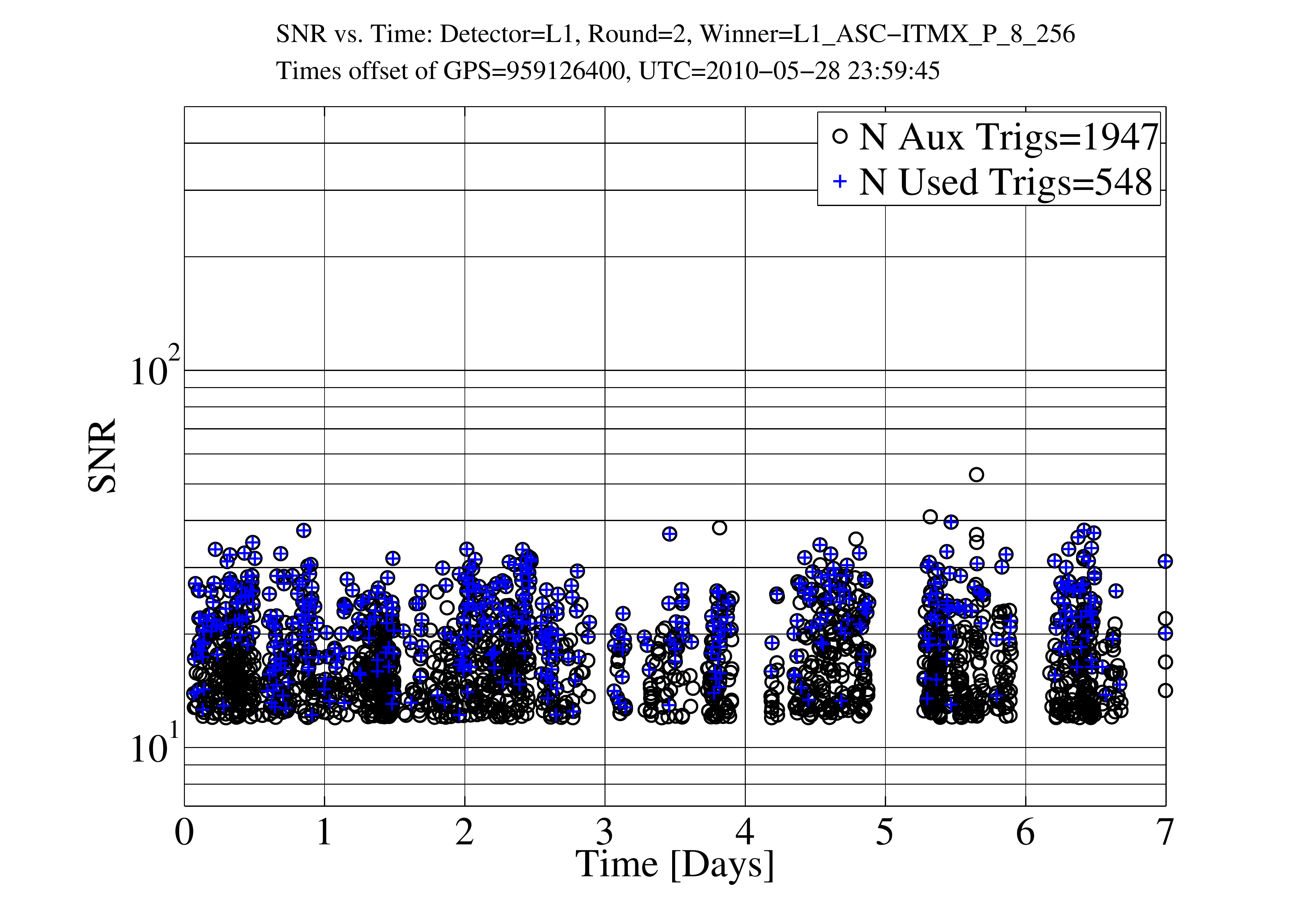}
		\caption{SNR versus time for noise transients in the ITMX pitch channel. The subset of these coincident with a transient in \emph{h(t)} are marked with blue plus signs.}
		\label{fig:snr-time-itmxp}
	\end{minipage}
	\hfill
	\begin{minipage}[t]{0.5\textwidth}
		\vspace{0pt}
		\includegraphics[width=1.2\linewidth]{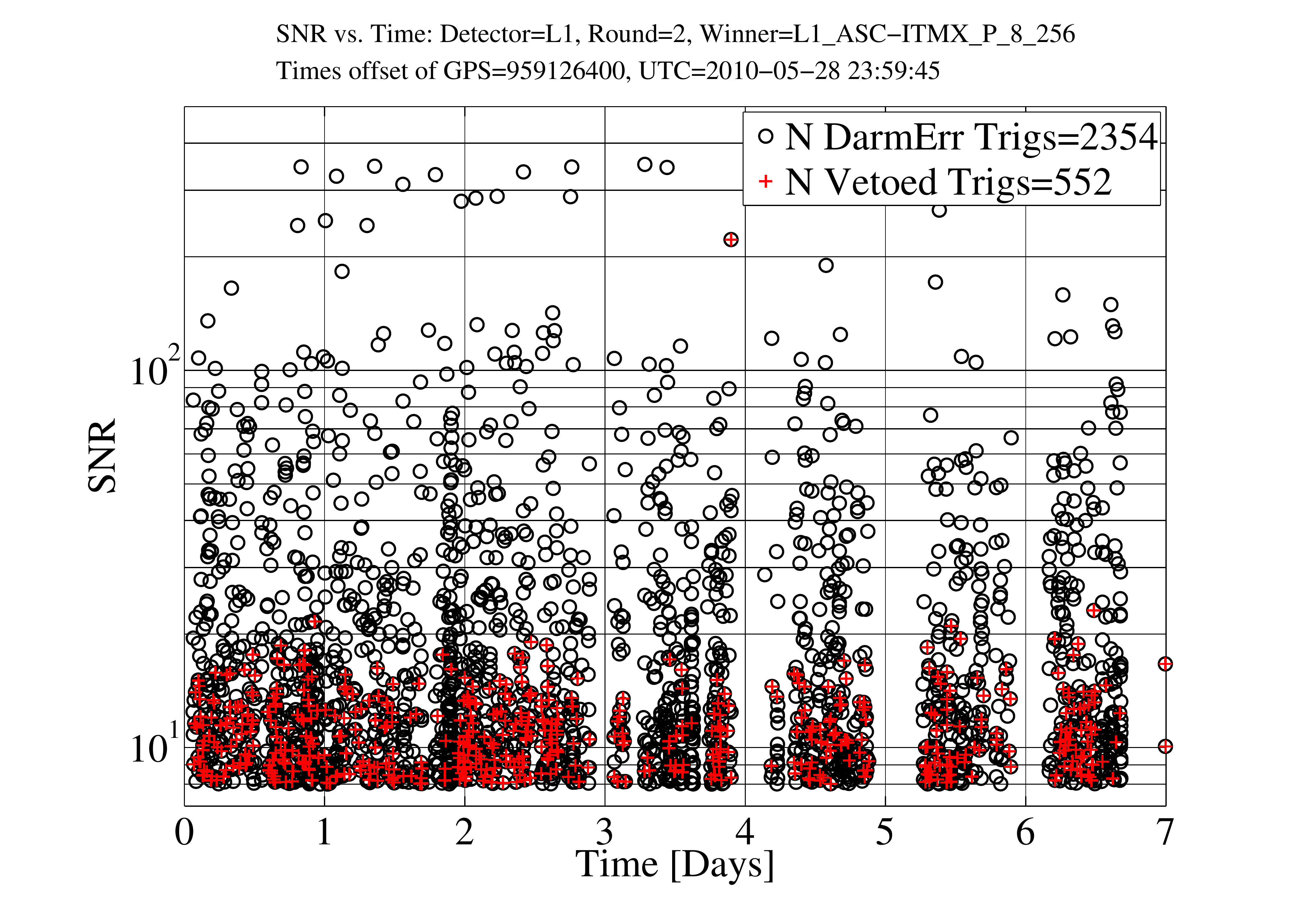}
		\caption{SNR versus time for noise transients in \emph{h(t)}. The subset of these vetoed by the ITMX pitch channel are marked with red plus signs.}
		\label{fig:snr-time-do2}
	\end{minipage}
\end{figure}

The second round was won by the channel \verb|ASC-ITMX_P|, hereafter \emph{ITMX Pitch}, which represents the pitch angular motion of the input-coupler optic for the long Fabry-Perot Cavity in the X-arm of the interferometer. Figure~\ref{fig:snr-time-itmxp} shows the SNR of the triggers in this channel versus time. Also indicated is the subset of triggers, 548 of the total 1947, that were coincident with triggers in \emph{h(t)}. The complete set of triggers were used to construct veto segments. Figure~\ref{fig:snr-time-do2} shows the SNR of the triggers in \emph{h(t)} versus time, and the subset, 552 of 2354, of these that were vetoed by the ITMX Pitch veto. This veto had a high 23\% efficiency and low 0.052\% deadtime.

\begin{figure}[h]
\includegraphics[scale=0.23, angle=-90]{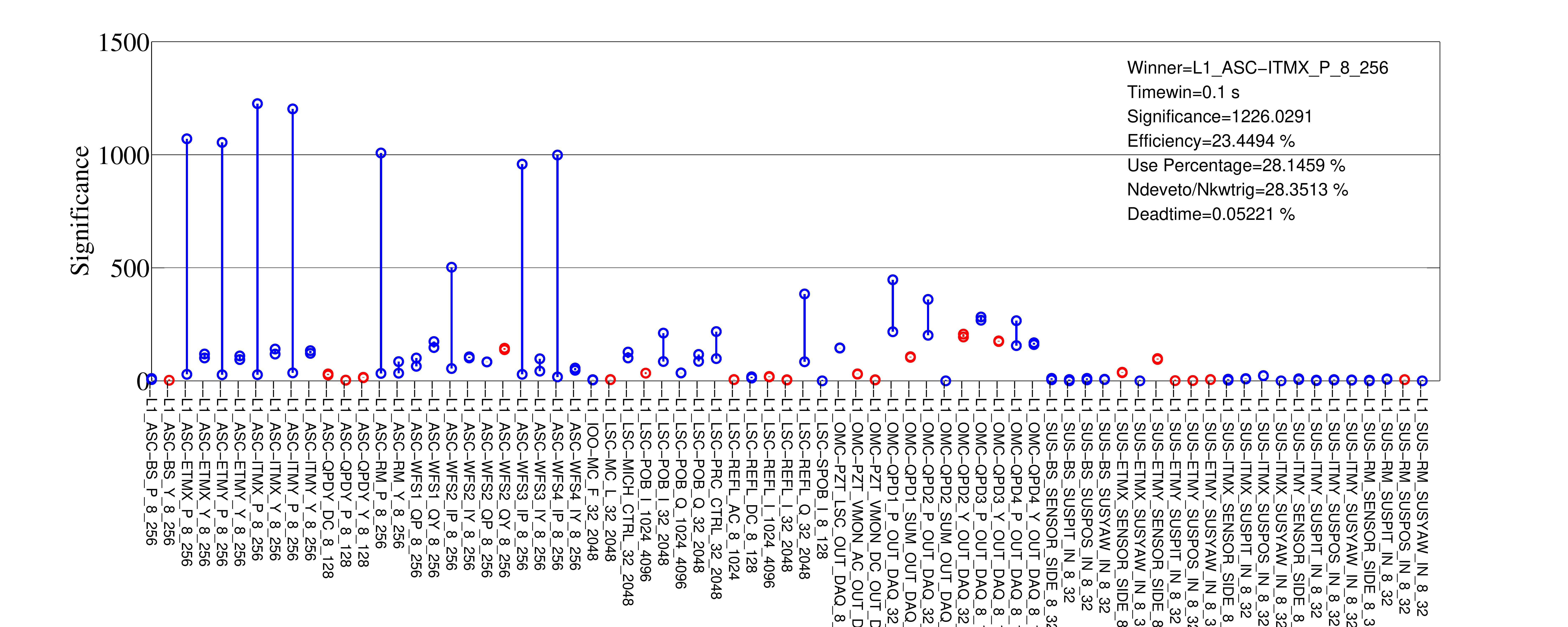}
\caption{Significance drop plot for the second round, won by ITMX Pitch. Each vertical line has two endpoints corresponding to the highest significance for its corresponding channel in a given round, and the following round, respectively. Blue lines indicate that the significance of that channel is lower in the later round, and red lines indicate the significance is the same or higher. A subset of instrumental sensing and control channels are shown. From left to right channel prefixes and their general description are ASC: alignment, IOO: frequency, LSC: length, OMC: output mode cleaner and SUS: suspensions.}
\label{fig:drop-itmxp}
\end{figure}

A novel feature of hveto is its ``significance drop plot''. This is useful for indicating whether any other channels than the round winner were sensitive to a population of triggers vetoed in a given round. It is produced by plotting the highest significance for each channel over all choices of threshold and time window in a given round as one end point of a vertical line, and then plotting the highest significance value for the following round as the other end point. A drop in significance for a channel from one round to the next indicates that some of its coincidences are the same as those vetoed by the round winner. If the significance stays the same (or increases, which can happen on occasion when the veto leads to a small decrease in overall analysis time), the channel has no relationship with the disturbance vetoed by the round winner. Typically the winning channel for a given round will have a sharp decrease in significance, indicated by a tall line. Note however that the significance of a winning channel does not always drop identically to zero in the following round, because the same channel may still have some coincidences at a lower SNR and/or wider time window. 

Significance drop plots are useful for identifying ``families" of channels that are all sensitive to the same type of disturbance. This can provide more information about the true origin of a disturbance, for example by localizing it to channels associated with a given building or subsystem. Figure~\ref{fig:drop-itmxp} shows the drop plot for the round won by ITMX Pitch. The first 24 channels from left to right across the x-axis correspond to alignment sensing and control systems. The largest significance drop is for the round winner, but the next seven largest significance drops are all for other channels in the interferometer that sense pitch alignment. The corresponding channels for yaw motion (for example \verb|ASC-ITMX_Y|) show little or no significance drop. This indicates that the underlying disturbance was nearly entirely in pitch, and was sensed throughout the interferometer.

Note that a non-hierarchical veto method based on significance would apply vetoes from eight different pitch alignment sensors here, each with imperfect use percentage and therefore additional deadtime, before applying a veto related to a different disturbance. In hveto, once the ITMX Pitch veto segments are applied, the significance of all pitch-related alignment channels drops considerably and none is selected again until ITMY Pitch in the tenth round.      

\begin{figure}[ht]
	\begin{minipage}[t]{0.5\textwidth}
		\vspace{0pt}
		\includegraphics[width=1.2\linewidth]{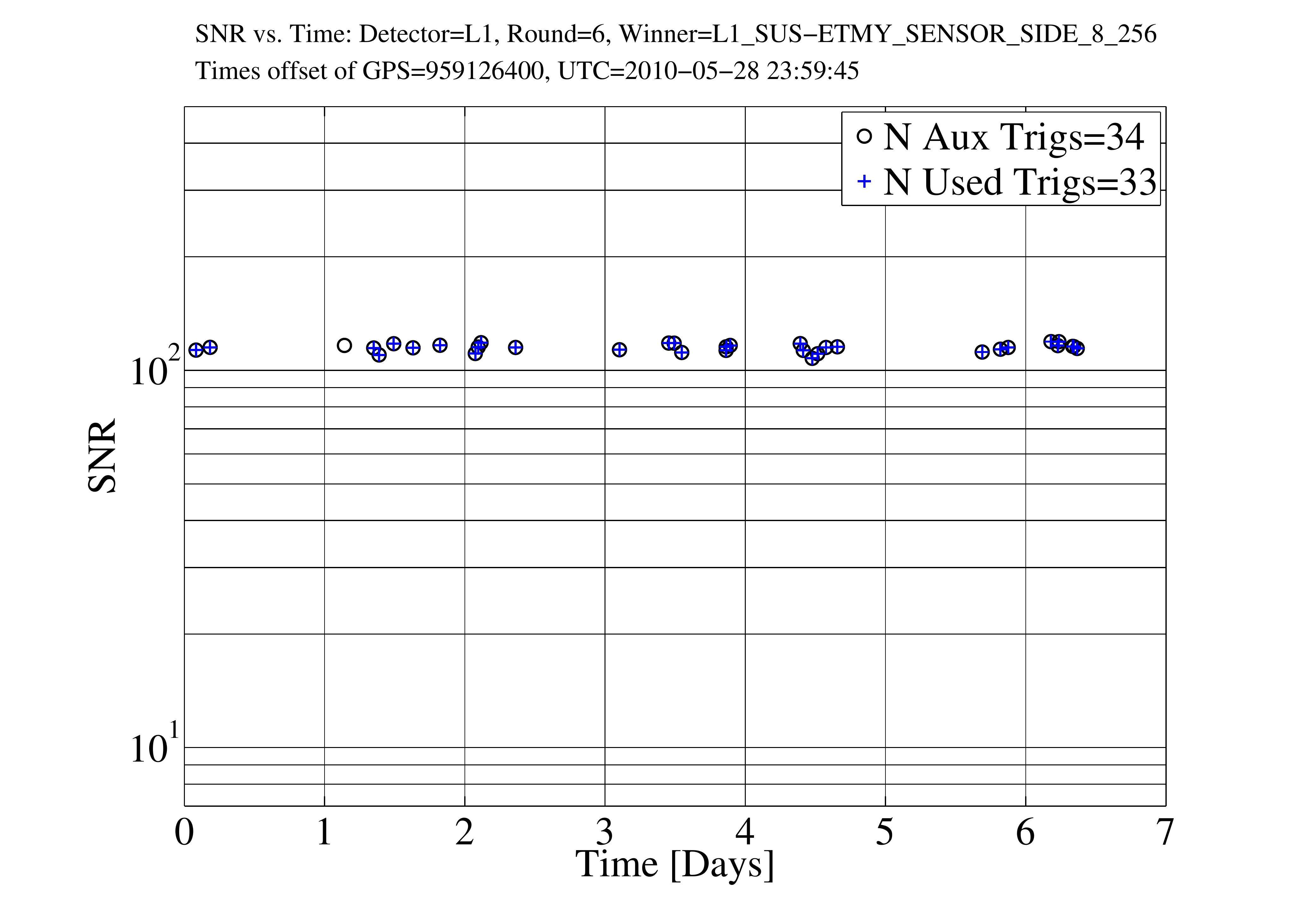}
		\caption{SNR versus time for noise transients in the Side Sensor channel. The subset of these coincident with a transient in \emph{h(t)} are marked with blue plus signs.}
		\label{fig:snr-time-side}
	\end{minipage}
	\hfill
	\begin{minipage}[t]{0.5\textwidth}
		\vspace{0pt}
		\includegraphics[width=1.2\linewidth]{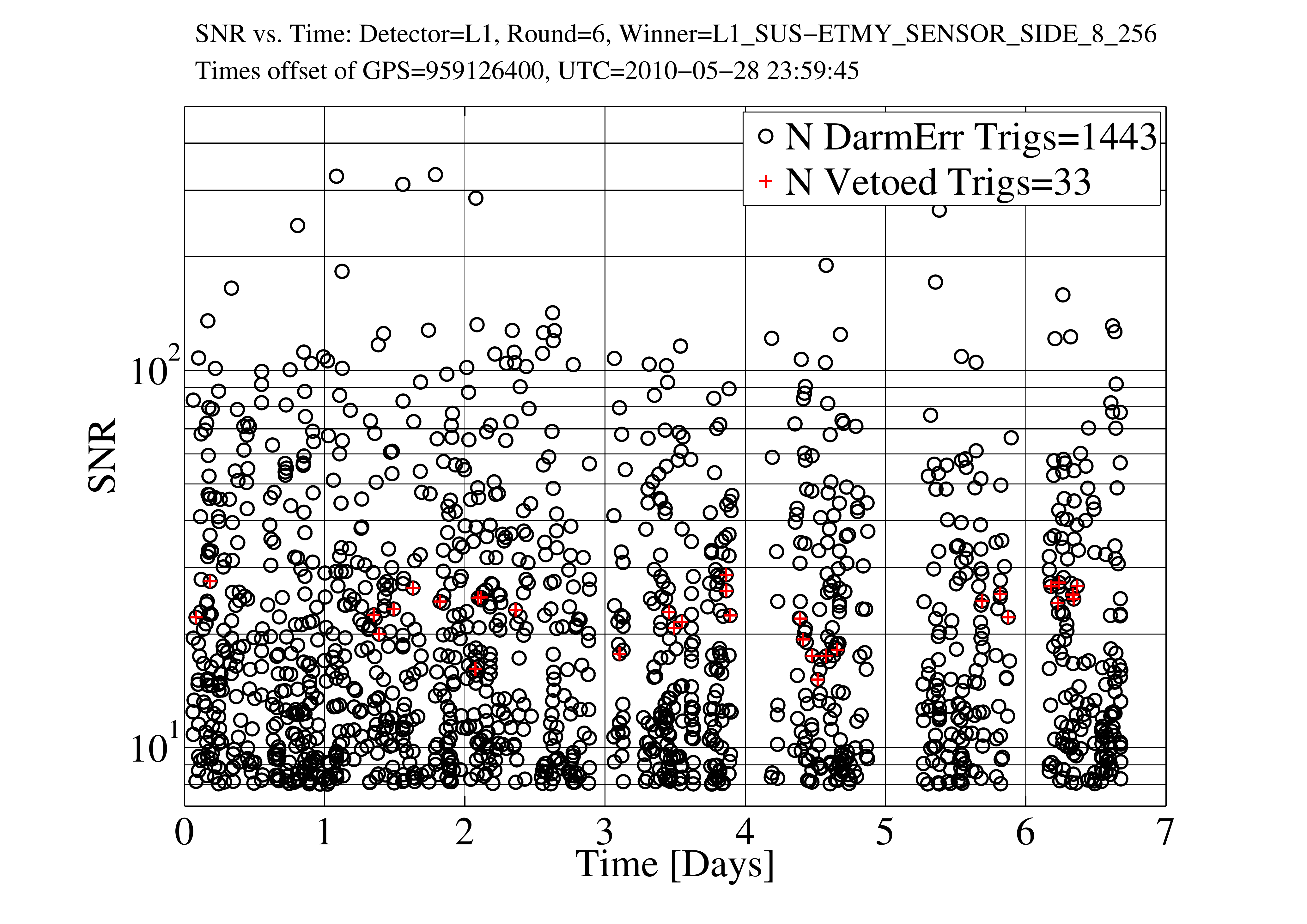}
		\caption{SNR versus time for noise transients in \emph{h(t)}. The subset of these vetoed by the Side Sensor channel are marked with red plus signs.}
		\label{fig:snr-time-do6}
	\end{minipage}
\end{figure}

As a second example, the sixth round was won by the channel \verb|SUS-ETMY_SENSOR_SIDE|, hereafter \emph{Side Sensor}, an optical monitor of the side-to-side position motion of the end test mass suspension of the Y-arm. Figure~\ref{fig:snr-time-side} shows the SNR of the triggers in the Side Sensor channel versus time, and the subset of these triggers that coincided with a trigger in \emph{h(t)}. The glitches, which were thought to have been due to a digital issue, repeated throughout the week with an SNR of just above 100. Of the 34 total triggers at this SNR threshold, 33 vetoed a trigger in \emph{h(t)}, giving a use percentage of 97\%, indicating a highly selective veto. Figure~\ref{fig:snr-time-do6} shows the SNR of the triggers in \emph{h(t)} versus time, and the subset that are vetoed by the Side Sensor channel. The combined information in these figures suggests that a population of noise transients with SNR of around 100 in the Side Sensor correspond to triggers in \emph{h(t)} with an SNR of 20-30. The fact that the transients in the Side Sensor have larger SNR than the corresponding transients in \emph{h(t)} is consistent with a causal coupling from the Side Sensor.

\begin{figure}[h]
\includegraphics[scale=0.23, angle=-90]{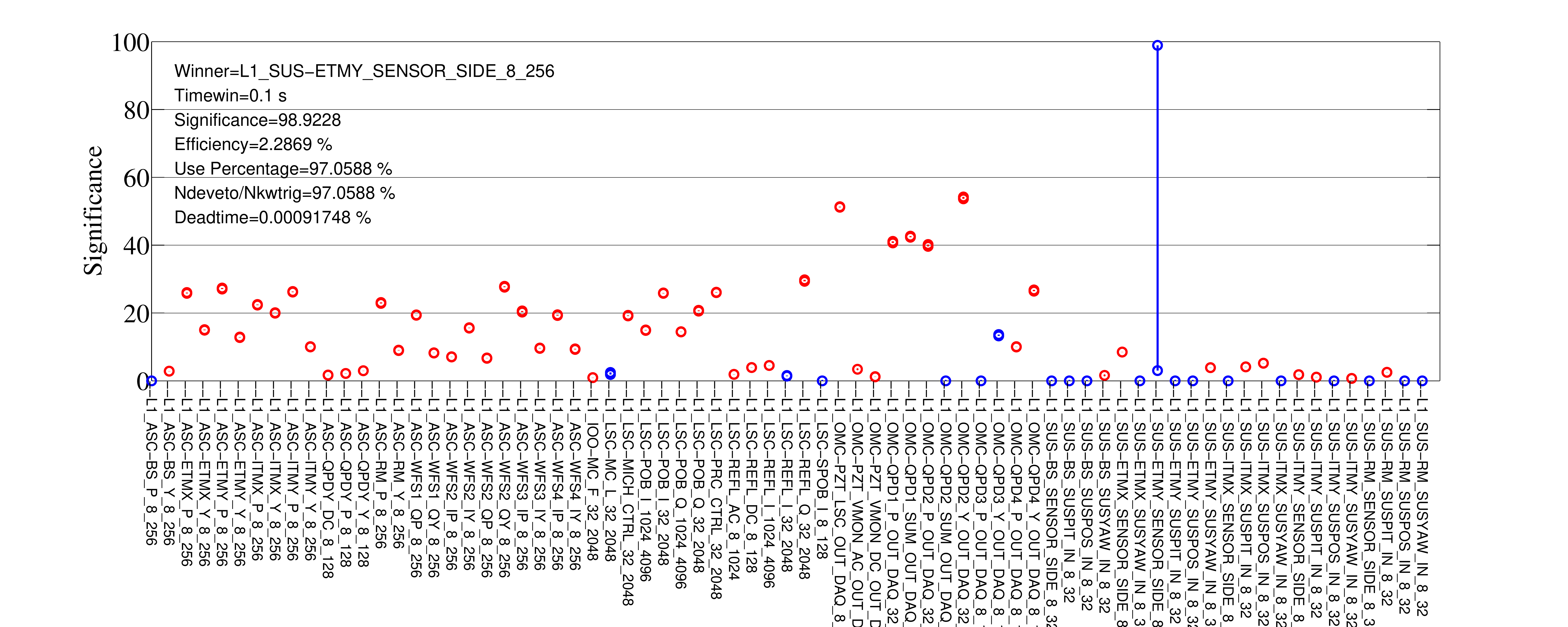}
\caption{Significance drop plot for the sixth round, won by the Side Sensor. Here no other channels show a relationship with the vetoed disturbance.}
\label{fig:drop-side}
\end{figure}

Figure~\ref{fig:drop-side} shows the drop plot for the round won by the Side Sensor. In contrast to the previous example, the Side Sensor is the only channel that experiences a large significance drop. This indicates that the Side Sensor was acting alone - the population of glitches it sensed were seen by no other auxiliary channels. 

\begin{figure}[ht]
	\begin{minipage}[t]{0.5\textwidth}
		\vspace{0pt}
		\includegraphics[width=1.2\linewidth]{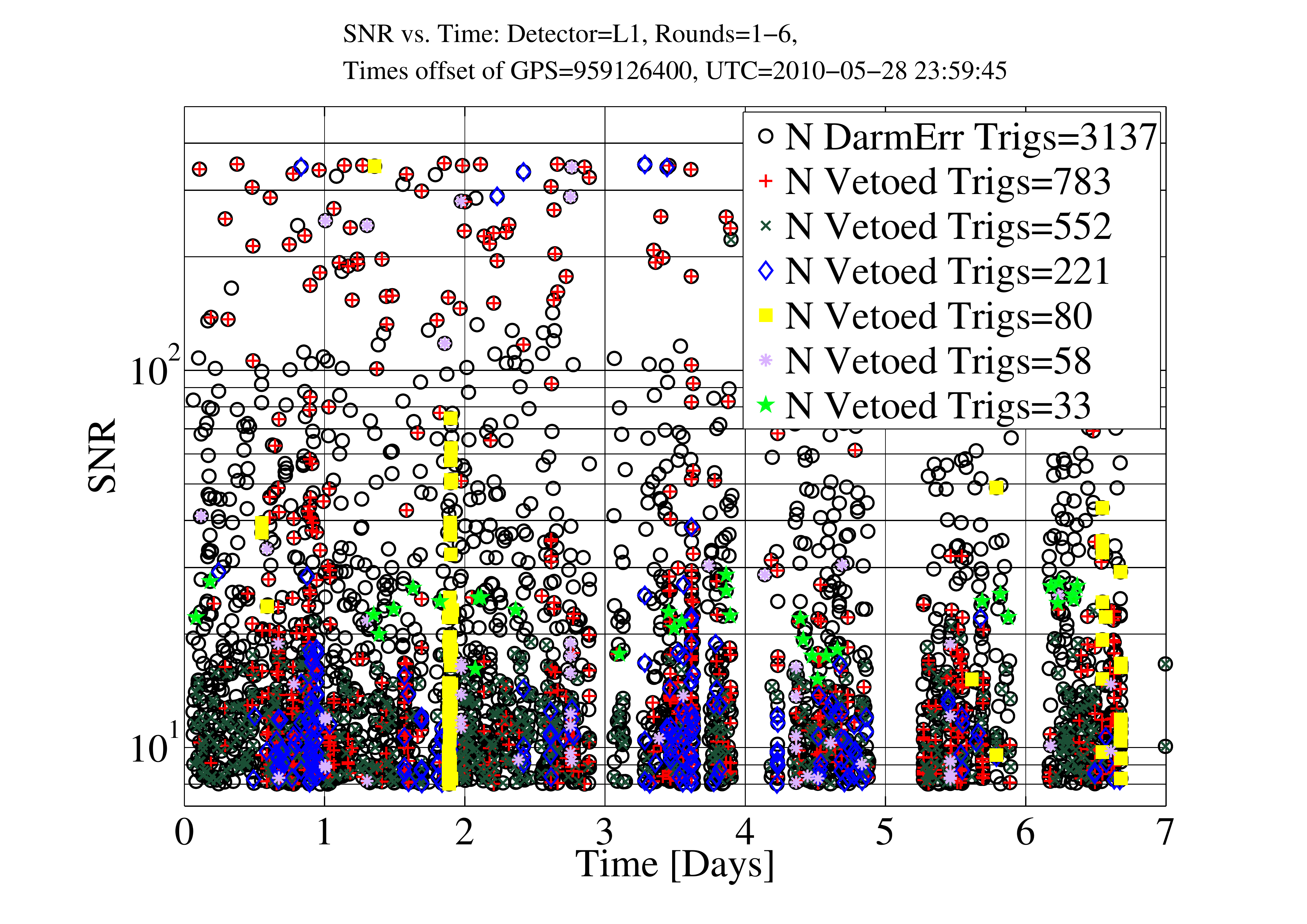}
		\caption{SNR versus time for noise transients in \emph{h(t)} (black), and the subset of these vetoed in each of the first six rounds (colored markers).}
		\label{fig:six-rounds}
	\end{minipage}
	\hfill
	\begin{minipage}[t]{0.5\textwidth}
		\vspace{0pt}
		\includegraphics[width=1.2\linewidth]{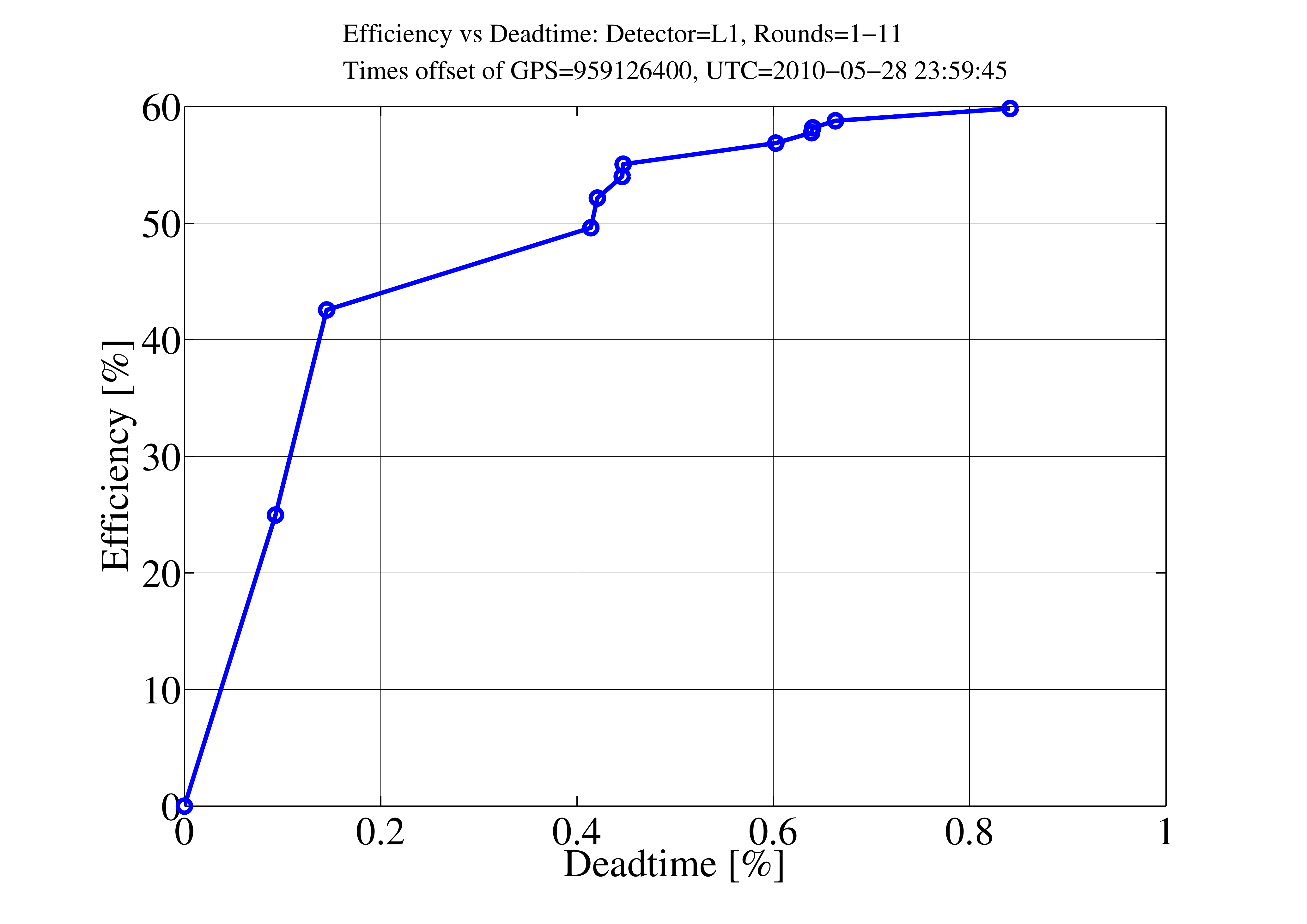}
		\caption{Efficiency versus deadtime for all rounds. Each round is marked with a circle (the initial condition of zero deadtime and efficiency is also marked). The slope of each line segment is the efficiency to deadtime ratio for that round.}
		\label{fig:effdt}
	\end{minipage}
\end{figure}

The previous two examples illustrate the ability of hveto to distinguish between different classes of disturbances, by identifying the families of channels that sense them and the degree to which they couple to \emph{h(t)}. Figure~\ref{fig:six-rounds} gives an indication of the variety of different glitch classes that hveto can identify. Here the starting set of \emph{h(t)} triggers are shown with black circles, and the vetoed triggers through the first six rounds are shown with colored markers.  Qualitatively the disturbances range from well-localized in time or in SNR to more isotropically distributed.   

Finally, an overview of the effectiveness of all vetoes can be seen in a plot of the cumulative efficiency versus cumulative deadtime, as shown in Figure~\ref{fig:effdt}. During this week 60\% of \emph{h(t)} triggers are vetoed at a cost of only 0.85\% deadtime, for a ratio 71. Note that the efficiency to deadtime ratio for each round is always much greater than one, indicating very effective vetoes. This ratio decreases overall as the rounds increase. However the slope for each successive round does not always decrease, because although the statistical significance is related to the efficiency to deadtime ratio, they are not directly proportional. 


\section{Summary and discussion}\label{sec:sum}

The algorithm described here identifies statistical relationships between putative gravitational-wave triggers in \emph{h(t)} and triggers in auxiliary channels that have no astrophysical sensitivity. It operates hierarchically to produce a minimal set of veto conditions that have a high efficiency to deadtime ratio. Once a veto condition is identified, its effects are removed, and new veto conditions are sought out using only the remaining triggers, revealing potentially different glitch mechanisms. In addition to vetoes, the algorithm produces information such as drop plots that can be used to identify families of channels that sense the same disturbances. 

Although this paper described illustrative results from only one detector, this algorithm has been used on all LIGO and Virgo detectors and those results will be presented in collaboration papers. It has proved useful for LIGO and Virgo science by producing vetoes that reduced the background in gravitational-wave analyses and providing hints used to develop data quality flags and to improve the detectors. This and complementary methods will reduce the impact of non-Gaussian noise on searches for gravitational waves with Advanced LIGO~\cite{aligo} and Advanced Virgo~\cite{avirgo}. 


\begin{ack}
The authors are grateful to the National Science Foundation for support under award numbers PHY-0970147, PHY-0854812, PHY-0653550 and PHY-0757957. The development of this method followed in part from an exploratory hierarchical veto implementation by Vijay Kaul (Maryland). We are grateful for ideas and encouragement from the gravitational-wave community and in particular our many colleagues in the LIGO Scientific Collaboration and the Virgo Collaboration.
\end{ack}


\end{document}